\documentclass[10pt, twocolumn, journal]{IEEEtran}

\usepackage{cite}

\usepackage[pdftex]{graphicx}
\usepackage[caption=false,font=footnotesize]{subfig}

\usepackage{amsmath}

\usepackage{url}

\usepackage{algpseudocode}

\usepackage[cmintegrals]{newtxmath}

\usepackage{lipsum}

\usepackage{siunitx}

\usepackage{soul,color}

\usepackage{mathrsfs}

\usepackage{array}

\usepackage[inline]{enumitem}

\usepackage{hyperref}

\usepackage{multirow}
\usepackage{array}
\newcolumntype{L}[1]{>{\raggedright\let\newline\\\arraybackslash\hspace{0pt}}m{#1}}
\newcolumntype{C}[1]{>{\centering\let\newline\\\arraybackslash\hspace{0pt}}m{#1}}
\newcolumntype{R}[1]{>{\raggedleft\let\newline\\\arraybackslash\hspace{0pt}}m{#1}}

\usepackage{todonotes}
\definecolor{morange}{rgb}{0.5,0.2,0.1}
\definecolor{mblue}{rgb}{0,0.1,1.0}
\definecolor{mred}{rgb}{1,0,0}
\definecolor{mgreen}{rgb}{0.2,0.4,0}

\newcommand{\treview}[1]{{#1}}

\begin{document}
%
\title{Full Duplex Integrated Access and Backhaul for \\5G NR: Analyses and Prototype Measurements}


 

\author{Gee Yong Suk,~\IEEEmembership{Student Member,~IEEE}, Soo-Min Kim,~\IEEEmembership{Student Member,~IEEE}, Jongwoo Kwak,~\IEEEmembership{Student Member,~IEEE}, Seop Hur, Eunyong Kim, and
Chan-Byoung Chae,~\IEEEmembership{Fellow,~IEEE}

\thanks{G. Y. Suk, S.-M. Kim, J. Kwak and C.-B. Chae are with the School of Integrated Technology, Yonsei University, Korea (E-mail: \{gysuk, sm.kim, kjw8216, cbchae\}@yonsei.ac.kr); S. Hur and E. Kim are with Samsung Electronics Co., Ltd., Korea (E-mail: \{s.hur, eunyong.kim\}@samsung.com).}
}


\maketitle

\begin{abstract}
Researchers for the third-generation partnership project (3GPP) have been exploring---as a cost-effective alternative to wired backhaul---integrated access and backhaul (IAB) frameworks for 5G new radio (NR). A promising solution for this framework is the integration of full duplex (FD) technologies to enhance the spectral efficiency and efficiently utilize network resources. This approach, which we refer to as FD IAB, involves a significant technical challenge---self-interference (SI) in the IAB framework. In fact, this challenge casts doubt over the performance and feasibility of FD IAB. In this article, we introduce the FD IAB framework and its enabling technologies and also evaluate the framework's link-level SI reduction and system-level downlink throughput performance. Thereafter, we validate the attenuation level in the SI channel with antenna separation and high directional gain through 28~GHz hardware prototype measurements. Our numerical evaluations and hardware prototype measurements confirm that FD IAB represents a promising framework for 5G NR.
\end{abstract}


\section{Introduction}
To realize the ambitious visions pertaining to future 5G networks, the third-generation partnership project (3GPP) has completed the standardization of a new radio (NR) access technology called 5G NR~\cite{3GPP2018NR}. One of the distinctive features of this standardization is its use of the millimeter-wave (mmWave) frequency band. The large available spectrum of the mmWave band enables a significant enhancement \treview{in} transmission speeds. However, owing to the severe path loss and penetration loss experienced therein, the system suffers from limited coverage and capacity. 

To overcome the path and penetration loss issues in the mmWave frequency band, engineers have exploited analog beamforming by focusing the signal power into narrow beams~\cite{Samsung2014mmWave}. The cooperative operation of multiple antenna elements enables the formation of a highly directive beam. Thus, the \treview{short} wavelength of the \treview{mmWave} band may also be attributed to analog beamforming, which enables a large number of antenna elements to fit into a compact form factor. In general, narrow beams not only increase the propagation-path length but also reduce the interference among various links, thereby providing significant potential for spatial multiplexing gains. 

Another common approach for enabling extended cell coverage and capacity expansion is network densification. The objective here is to provide a reliable access channel by reducing the inter-site distance, that is, deploying more cellular base stations (BSs) with \treview{a} smaller coverage in a given area~\cite{lopez2015towards}. However, a significant drawback to this approach is that the large number of BSs and the consequent fiber backhaul interconnections incur substantial capital/operational expenditures (CAPEX/OPEX). 

Nevertheless, future networks are expected to be highly dense to support the high standards of future applications, such as virtual/augmented reality, the Internet of Things, edge computing, vehicle-to-everything. However, traditional fiber-backhauling is often an economically impractical solution for carrier operators. In this context, integrated access and backhaul (IAB) technology has emerged as a cost-effective alternative to the traditional fiber-backhauled system. In the case of IAB, only a \treview{few} of the BSs are connected to the traditional wired infrastructures while the other BSs relay the backhaul traffic wirelessly~\cite{polese2020integrated, 3GPP2018SI}. 
\begin{figure*}[t]
	\begin{center}
		{\includegraphics[width=2.0\columnwidth,keepaspectratio]
			{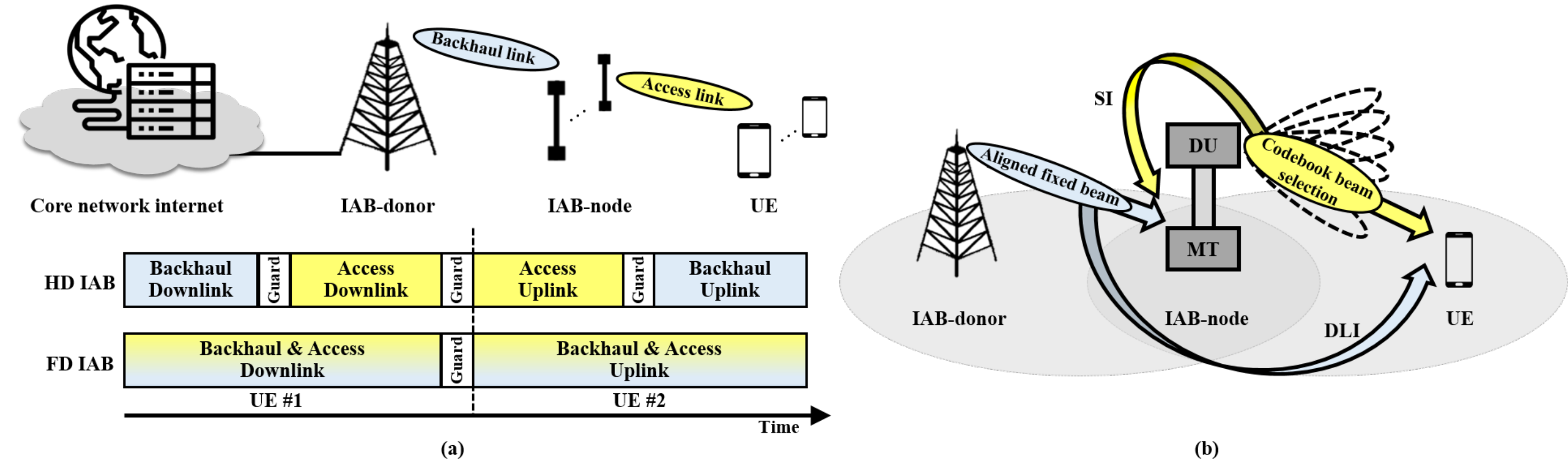}%
			\label{fig_polar_plot_25}}
		\caption{(a) IAB system architecture with time-domain partitioning; (b) functional structure of IAB-node in downlink and sources of interfering signals.}
		\label{fig_system_model}
	\end{center}
\end{figure*} 
In a typical IAB framework, the access and backhaul links share the same frequency spectrum, which results in the resource collision problem; thus, resource management is required to resolve this issue. Owing to the simplicity of implementation, many previous studies have incorporated half duplex (HD) constraints in their frameworks~\cite{polese2020integrated, 3GPP2018SI}, which we refer to as HD IAB. In HD IAB, the access and backhaul links must use the given radio resources orthogonally, be it time or frequency. While this helps prevent collisions between the two separate links, it fails to exploit the full potential of the given radio resources.

\treview{In contrast, a smarter IAB framework with full duplex (FD) techniques, which we denote as FD IAB, may simply rule out the HD constraint.} FD is an advanced technology, the objective of which is to realize more efficient utilization of the given radio resources by allowing transmission and reception to occur at the same time and in the same frequency resource block. Theoretically, FD technology not only doubles the spectral efficiency but also grants more flexibility in the design of wireless protocols~\cite{Sachin2013Full, Sabharwal2014IBFD}.   
The application of FD techniques to the inter-relaying node has been previously studied based on the concept of a FD relay. 
\treview{However, the majority of the previous analyses and experiments were focused on} the sub-\SI{6}{GHz} frequency band and transmission powers typically much lower than that of the next generation Node B (gNB), e.g., Wi-Fi~\cite{GLiu2015FDRelay,kim2015survey}. \treview{Moreover, practical studies need to be conducted to investigate the applicability of FD relays to the IAB of 5G NR. This is because the significant self-interference (SI) resulting from the infrastructure casts doubt on the performance and feasibility of the system.} 

In this study, we first introduce the FD IAB framework and discuss its potential. Then, we analyze its performance and feasibility. Our analysis consists of link-level and system-level performance evaluations based on realistic IAB configuration, and a feasibility test through \SI{28}{\giga\hertz} hardware (H/W) prototype measurements. The remainder of this article is organized as follows. In Section~\ref{sec.2FD_IAB}, we present a tutorial \treview{on} the concept and preliminaries of FD IAB. In Section~\ref{sec.3Performance}, we present numerical performance analyses of FD IAB at the link and system levels. In Section~\ref{sec.4P_Performance}, we present some H/W prototype measurement results to validate the feasibility of FD IAB. In Section~\ref{sec.conclusion}, we discuss some remaining technical challenges and present the conclusion of this study.


\begin{figure*}[t]
	\begin{center}
		{\includegraphics[width=2.0\columnwidth,keepaspectratio]
			{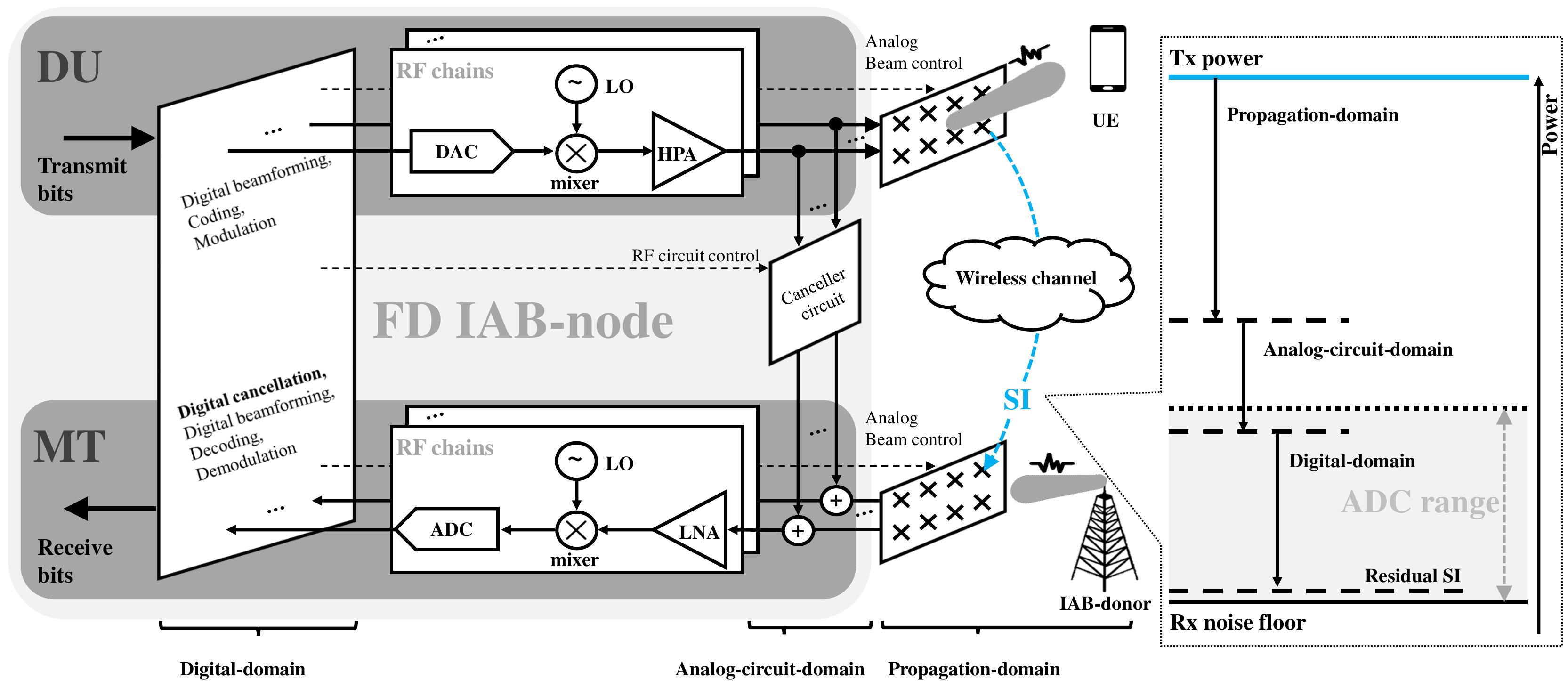}}
		\caption{Block diagram of the SI reduction procedure in a FD IAB-node in downlink scenario. The illustration is partially adopted from~\cite{Sabharwal2014IBFD}.}
		\label{Fig.simulModel}
	\end{center}
\end{figure*}

\section{{\fontsize{11}{14}\selectfont FD IAB: A Concept and Preliminaries}}
\label{sec.2FD_IAB}
\subsection{Full-Duplex IAB Framework}
Included in the 3GPP NR Release 16, the work item on IAB was finalized in 2020~\cite{3GPP2018SI}. 
A typical IAB architecture is illustrated in Fig.\ref{fig_system_model}, where only some of the gNBs (IAB-donors) are connected to the traditional wired infrastructures (core network internet) via fiber, while the rest of the gNBs (IAB-nodes) relay the backhaul traffic wirelessly. The typical scenario involves an in-band system in which the backhaul and access links share the same frequency spectrum and the IAB-nodes decode and forward the backhaul traffic. In this architecture, each IAB-node possesses two NR functional blocks, that is, a mobile termination (MT) and a distributed unit (DU). The MT maintains the wireless backhaul link with an upstream IAB-node or IAB-donor, while the DU provides the access link with the downstream IAB-nodes or user equipment (UE). Commonly performed at the IAB-donor and the distributed IAB-nodes are the lower-layer functions, such as radio link control (RLC), medium access control (MAC), physical (PHY), and backhaul adaptation protocol (BAP). In contrast, functions performed at the central unit (CU) of the IAB-donor are upper-layer ones, such as radio resource control (RRC) and packet data convergence (PDC).

As noted above, an interference problem occurs owing to the spectrum sharing of the backhaul and access links. To address this issue, HD constraints, such as time-division multiple access (TDMA), illustrated in Fig.~\ref{fig_system_model}(a), frequency-division multiple access (FDMA), or spatial division multiple access (SDMA) are incorporated into the IAB system. The wide adoption of HD constraints from previous researchers may be largely attributed to the relative simplicity of their system design and implementation~\cite{polese2020integrated,3GPP2018SI}.



In contrast, FD IAB, with its incorporation of an in-band FD relay, offers an efficient and flexible system design. An FD IAB is a network framework in which a relay-node carries---at the same time and \treview{in} the same frequency resource block---the reception and transmission, thereby obviating the need for HD constraints. 
\treview{Considering the guard intervals between the backhaul and access connection,} as illustrated in Fig.~\ref{fig_system_model}(a), FD IAB framework has the potential to achieve a spectral efficiency more than double that of HD IAB. Furthermore, the application of FD technology can reduce end-to-end and feedback delays and improve wireless protocol design~\cite{kim2015survey}. 

However, there exists a critical technical challenge regarding SI that must be addressed. As illustrated in the downlink example in Fig.~\ref{fig_system_model}(b), SI is an undesired interfering signal that is transmitted from the DU of an IAB-node to the UE and is received by the MT of the node (in the uplink case, the MT becomes the self-interfering source). Owing to the MT's proximity to the interference source (DU) compared to the proximity to the desired signal's source, SI is a powerful interference signal that can cause significant performance loss. Furthermore, because SI increases with the Tx power, deploying IAB-nodes with high Tx power requires state-of-the-art SI reduction performance. For example, the transmit power at the IAB-DU may be up to \SI{46}{dBm} in a macrocell scenario, which implies that an SI reduction more than \SI{120}{dB} is required to cut the SI close to the Rx noise floor of \SI{-90}{dBm}. Due to such strict performance requirements, the application of FD in infrastructure systems has been considered a challenge in the context of conventional FD radios~\cite{GLiu2015FDRelay,kim2015survey}.

Another issue \treview{to note} is the direct-link interference (DLI), which comprises \treview{an} interfering backhaul signal along a direct path to the UE of the access link, bypassing the relay node. 
Note that the access link of a DU and UE would be adaptively selected from its beam codebook, whereas the backhaul link of an MT and IAB-donor would have a stable backhaul link through a perfectly aligned fixed beam. Consequently, the SI experienced at the MT may vary. 

\subsection{Self-Interference Reduction Techniques}


In this section, with a focus on FD IAB, we briefly summarize the existing SI reduction techniques~\cite{Sabharwal2014IBFD,GLiu2015FDRelay,kim2015survey}. SI reduction techniques can be categorized according to their domain of realization, that is, the propagation-domain, analog-circuit-domain, and digital-domain. The overall procedure of SI reduction for a given downlink scenario is illustrated in Fig.~\ref{Fig.simulModel}. The SI signal first propagates with the Tx power, and then its power is reduced, \treview{through} consecutive applications of SI reduction techniques, to a value close to the Rx noise floor. This continuous procedure is analogous to a group of people trying to drain a vessel of water by ingestion---the more one person drinks, the less the next person has to. 

\begin{figure*}[t]
	\centerline{\resizebox{2\columnwidth}{!}{\includegraphics{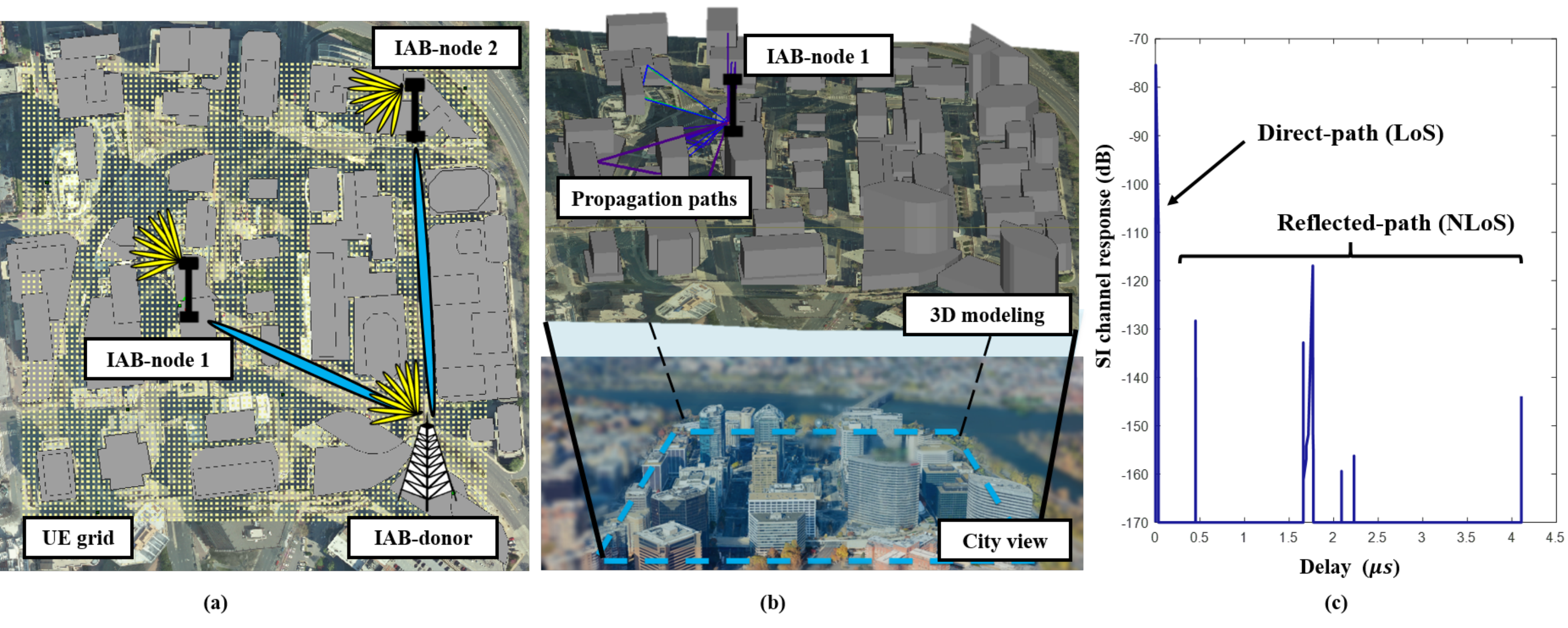}}}
	\caption{System-level simulation of IAB in 28~GHz 3D modeled environments: (a) Network deployment configuration; (b) 3D modeling of urban environment and realistic channel modeling via ray-tracing; (c) Example of the residual SI channel at IAB-node~1 ($d=\SI{0.1}{m}$).}	
	\label{Fig.systemLevel}
\end{figure*}

\subsubsection{Propagation-Domain SI Suppression}
SI suppression in the propagation-domain is likened to the first person drinking the water in the vessel. Before the SI signal reaches the MT, it is suppressed in the propagation-domain via electromagnetic isolation of the transmitting and receiving antennas. This isolation is performed by using a combination of path loss, antenna directionality, and cross-polarization~\cite{EEverett2014Passive}. 

A straightforward method of suppressing the SI is to make the SI signal experience more \textit{path loss}. This can be done either by physically separating the RF front-ends of the Tx and Rx by a large distance or by placing an electromagnetic barrier between them. \treview{Owing to the severe path loss experienced at the mmWave band, antenna separation is highly effective in FD IAB. In practice, though, to prevent the bulky form factor of the network equipment, engineers should consider the tradeoff between the amount of SI suppression and the physical size of the device.} 

Another approach is to exploit the \textit{antenna directionality} of the RF front-end. By employing directional transmit and receive antennas that focus their radiating/sensing capability in specific directions, the interference between them can be avoided in advance. Because the design and control of highly directive analog beams are readily accessible by manipulating the large antenna array\footnote{In case of an orthogonal frequency division multiplexing (OFDM) system, analog beamforming control will take place in the time-domain.} in the mmWave system, this approach will be particularly effective for FD IAB. 

Another notable method for propagation-domain SI suppression is \textit{cross-polarization}. The occurrence of interference is prevented in advance via the adoption, at the RF front-ends, of a set of cross-polarized antennas, which only transmit/receive horizontally/vertically polarized signals~\cite{prototype}. 


\subsubsection{Analog-Circuit-Domain SI Cancellation}
The next SI reduction occurs in the analog-circuit-domain. As shown in Fig.~\ref{Fig.simulModel}, the canceller circuit taps the outgoing transmit signal, regenerates an estimate of the SI signal, and subtracts it from the received signal\footnote{Unlike the structure presented in Fig.~\ref{Fig.simulModel}, which has a single high-cost high-power amplifier (HPA) per RF chain, an architecture with multiple low-cost lower-output PAs (a PA per antenna)~\cite{roberts2021millimeter} provides an economic benefit. In this case, tapping the signal before power splitting and multiple PAs makes more sense to reduce circuit complexity. However, because of its tapping point, this architecture will not be able to capture and cancel out the nonlinearities of the PAs through RF circuitry (analog-circuit-domain).}. To regenerate the SI signal, there are two approaches---the \textit{channel-aware} and \textit{channel-unaware} approaches. In the first approach, the knowledge of the SI channel is periodically acquired and the canceller circuit is actively tuned to mimic it. In the second approach, the canceller circuit is either tuned once, when the system is calibrated, or blindly adjust its tuning in the direction of reducing the residual SI measured at the Rx. A tradeoff between the training overhead and the accuracy of the regenerated signal exists across the different approaches.


\treview{In practice, the SIC in the analog-circuit-domain entails costs beyond training overhead such as additional power consumption and circuit complexity. Furthermore, for the adaptive circuit to support multi-stream or wideband signals, greater costs would be incurred. Thus, it is crucial to design a cost-efficient circuit structure and tuning algorithm.}

\subsubsection{Digital-Domain SI Reduction}
After passing through the propagation-domain and analog-circuit-domain, the SI that remains is then subjected, as depicted in Fig.~\ref{Fig.simulModel}, to digital-domain SI reduction---analogous to the last person drinking the water from the vessel. The most significant advantage of the operations in the digital-domain is the convenient implementation of complicated signal processing. However, digital-domain SI reduction is limited by the dynamic range of an analog-to-digital converter (ADC), as SI beyond the dynamic range will cause Rx saturation. Thus, to guarantee the SI reduction performance in the digital-domain, the magnitude of the residual SI power after the two previous domains should be lower than the magnitude by which the ADC's dynamic range is greater than the noise floor. In particular, the residual SI power following the two previous domains should reach the gray zone in the graph on the right-hand side of Fig.~\ref{Fig.simulModel}.

The conventional approach of digital-domain SI reduction is \textit{digital SI cancellation (SIC)}. The expected baseband-equivalent SI signal is regenerated based on the estimated residual SI channel and then subtracted from the received baseband digital signal. 
\treview{In a multi-stream scenario, \textit{digital precoding/combining} (\textit{or digital beamforming}) can also mitigate the cross-talk}\footnote{\treview{The cross-talk components are the SI signals induced by other adjacent transmitted data streams from the local MIMO (multiple-input multiple-output) Tx.}} \treview{components of the SI signal via digitally weighting complex-valued gains to each of the streams in an adaptive manner. The joint cooperation and its optimization between the analog/digital SIC and analog/digital beamfoming appears to be the emerging research topics~\cite{roberts2021millimeter}.
	
One of the main hurdle associated with the aforementioned methods are the nonlinearity caused by RF imperfections (e.g., nonlinearity of the power amplifier (PA), in-phase and quadrature (I/Q) imbalances, and phase noise at a local oscillator (LO)) and the computational complexity involved in addressing these issues. Efficient channel models and algorithms are therefore developed to consider these nonlinear characteristics}~\cite{kwack}.


\section{\fontsize{11}{14}\selectfont{Numerical Analyses and Discussions}}
\label{sec.3Performance}

In this section, we numerically analyze and discuss the performance of FD IAB for 5G NR. Interdependent simulations at the link-level and system-level are conducted.


\subsection{Network Deployment Configurations and SI Suppression}
Our analysis begins by setting the various network topologies in a virtual 3D environment. We consider an in-band IAB scenario with physically fixed relays, equipped with a single antenna and an RF chain at its DU and MT, over a \SI{120}{\mega\hertz} bandwidth with \SI{28}{\giga\hertz} center frequency. As illustrated in Fig.~\ref{Fig.systemLevel}(a), we employ one IAB-donor and two IAB-nodes to support the 4,527 equally spaced outdoor UEs, forming a rectangular grid in the \SI{500}{m}$\times$\SI{500}{m} region via maximum single-hop relaying. The IAB-donor and IAB-nodes are placed at the top of the buildings \SI{130}{m}, \SI{126}{m}, and \SI{99}{m} in height (approximately 33--43 floors). 

Shown in Fig.~\ref{Fig.systemLevel}(b) is a 3D urban model of the downtown area in Rosslyn City, Virginia, USA. 
To precisely compute the wireless propagation under given network configurations, we adopt ray-tracing algorithm, geometric optics, and the uniform theory of diffraction, using Wireless Insite. The computation reflects the impact of cell deployment, radiation pattern of the antenna, antenna separation, and reflection/diffraction from the surrounding environment on the SI suppression in propagation-domain. Fig.\ref{Fig.systemLevel}(c) illustrates a realization of a channel response of the residual SI. 

We assume that the network supports one UE at a given time instance. A simple scheduling that connects each UE to the cell providing the best access link in terms of the signal-to-noise ratio (SNR) is used~\cite{MS_schedule}. Similarly, for the DU to support its allocated UE, the maximum-SNR beam selection scheme is employed. For the access links thereof, the DUs share a universal codebook that divides 120$^{\circ}$ in the azimuth plane into eight main directions and 30$^{\circ}$ in the elevation plane into two main directions. The 16-beam codebook in the simulation is generated by mechanically rotating the radiation pattern of a directional antenna, which has a {20}~{dBi} directivity gain with an approximate {3}~{dB} beamwidth of 12$^{\circ}$. The MTs of the IAB-nodes employ a shared radiation pattern with the DUs of the IAB-donor and IAB-nodes. In contrast, the UEs employ a uniform dipole antenna with an isotropic radiation pattern in the azimuth plane. The DUs transmit their signals with a high transmission power of \SI{43}{dBm}.

\subsection{Link-Level Evaluation of SI Cancellation}

After acquiring the residual SI channel subsequent to the propagation-domain SI suppression, we conduct a link-level SIC simulation based on single-stream OFDM. A pilot-based, channel-aware, two-tap canceller circuit is considered for the analog-circuit-domain, and nonlinear digital SIC with the parallel Hammerstein model of the fifth order~\cite{kwack} is considered for digital-domain.

For the analog-circuit-domain SIC, we adopt a two-tap RF canceller. The fixed-delay values are pre-determined based on the delay of the direct path of each antenna separation case. For the cases of $d=\SI{2}{\meter}$, $d=\SI{1}{\meter}$, and $d=\SI{0.1}{\meter}$, the two delays of the canceller are set as $\{\SI{6}{\nano\second},\ \SI{8}{\nano\second}\}, \{\SI{3}{\nano\second},\ \SI{4}{\nano\second}\}$, and $\{\SI{0.3}{\nano\second},\ \SI{0.4}{\nano\second}\}$, respectively. 
Furthermore, the effect of nonlinear components in the actual RF circuit board, such as PA (\SI{20}{dB} gain and \SI{43}{dBm} $\text{P}_{\text{1dB}}$) and a 14-bit ADC, is considered. For the OFDM, the research team considered a fast Fourier transform size of 1024, a subcarrier number of 792, and cyclic prefix length of 140.
%

\begin{figure}[t]
	\centerline{\resizebox{1\columnwidth}{!}{\includegraphics{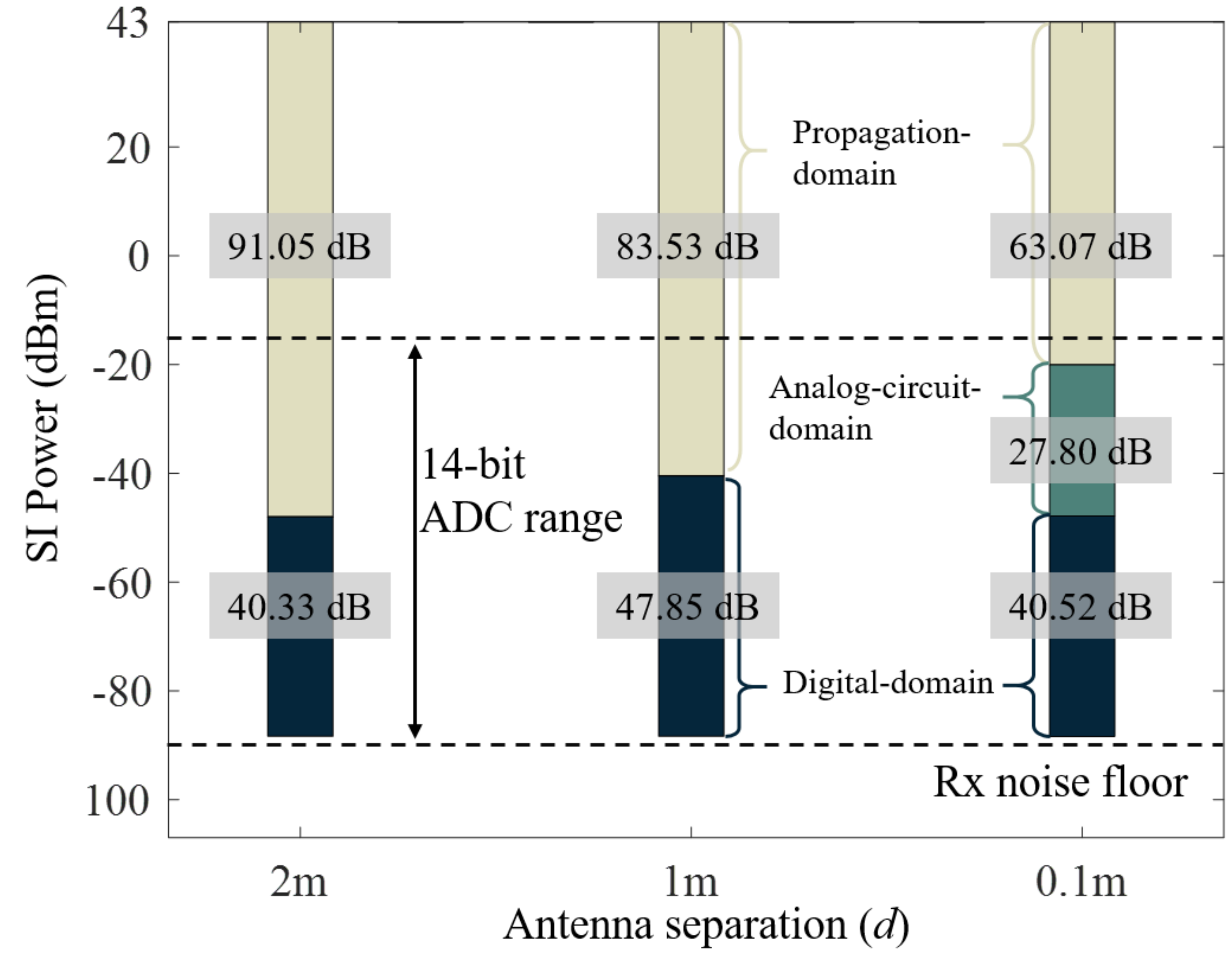}}}
	\caption{Simulated result for full SI reduction with different antenna separations.}
	\label{Fig.SIClink}
\end{figure}

Fig.~\ref{Fig.SIClink} presents the averaged SI reduction and its composition for all the transmission cases in the scenario with three different antenna separations. For all antenna separations, the average total SI suppression is sufficient to cut the residual SI to a value close to that of the Rx noise floor. For the $d=\SI{2}{\meter}$ and $d=\SI{1}{\meter}$ cases, the SI reduction in the propagation-domain alone managed to reach the effective ADC range (\SI{72.24}{dB} safely, with a 14-bit resolution~\cite{Sabharwal2014IBFD}) above the Rx noise floor; thus, the requirement for digital-domain SIC was satisfied. 
\treview{This observation leads to new opportunities wherein active-analog-domain SIC may be redundant for certain FD IAB configurations. This would enable a cost-efficient system without a canceller circuit.}
For the $d=\SI{0.1}{\meter}$ case, propagation-domain SI suppression was sufficient to be within the effective ADC range above the Rx noise floor on average. Yet, there were specific UE cases wherein the suppression was insufficient, which resulted in significant performance loss at the digital-domain cancellation. Therefore, analog-circuit-domain SIC have been applied to ensure that the residual SI reaches the ADC range. Overall, the impact of antenna separation and analog beamforming in the propagation-domain \treview{of the mmWave band} is substantial, constituting a significant proportion of the total SI reduction. 

\subsection{System-Level Evaluation of IAB Throughput}

In this section, we present the downlink throughput performance of various IAB configurations with antenna separation cases of $d=\SI{2}{m}$, $d=\SI{1}{m}$, and $d=\SI{0.1}{m}$. \treview{The downlink throughput of each UE is derived from the capacity of its access link, which is affected by DLI and the capacity of the backhaul link. For FD IAB, the capacity of the backhaul link is determined by the SI reduction performance.} 

We now observe the performance of FD IAB with full SI reduction (FD IAB), considering the previous result from the link-level simulation with full application of SI reduction techniques. As a benchmark, we also evaluate the network topology of a fibered-backhaul system with three wired gNBs, each positioned at the IAB-donor and IAB-node coordinates (fibered-backhaul); the same topology with HD constraints and operation (HD IAB); an ideal FD IAB that assumes zero SI (ideal FD IAB); and a FD IAB with only SI suppression in the propagation-domain (FD IAB without (w/o) full SI reduction). 

\begin{figure}[t]
	\centerline{\resizebox{1\columnwidth}{!}{\includegraphics{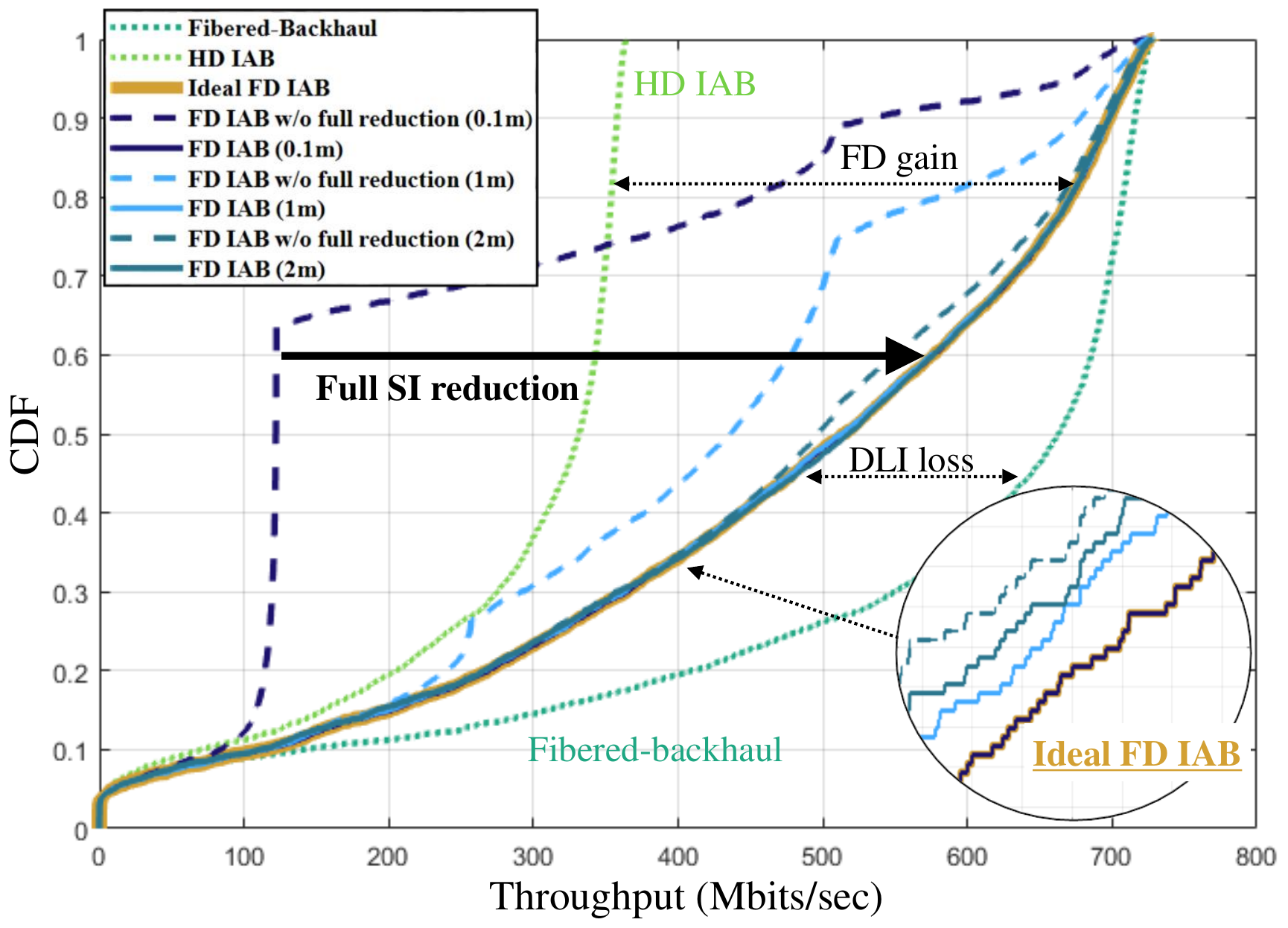}}}
	\caption{Downlink throughput cumulative distribution function (CDF) of outdoor UEs for different antenna separations.}
	\label{Fig.DTput}
\end{figure}

First, Fig.~\ref{Fig.DTput} shows that the FD IAB without full SI reduction suffers from performance degradation owing to the SI, which intensifies as the antenna separation decreases. Note that, in the case of $d=\SI{0.1}{m}$, the performance of FD IAB became worse than that of the HD IAB.
In the cases of $d=\SI{2}{m}$ and $d=\SI{1}{m}$ without full SI reduction, the degradation is evident but not as severe as that for $d=\SI{0.1}{m}$. This is because the backhaul link was sufficiently solid to withstand the residual SI for the two cases, which can be largely attributed to its perfectly aligned fixed beam between the IAB-donor and IAB-node. It should be noted that the SI deteriorates the quality of the backhaul link (see Fig.~\ref{fig_system_model}(b)), which forms a virtual upper bound for the access link. Combined with the discrete selection of adaptive modulation schemes at the backhaul, this bottleneck results in an irregular curve for the FD IAB without full reduction.

In contrast, the ideal FD IAB curve represents the upper bound of the performance of FD IAB systems. Note that the performances of all the considered FD IABs are approximately the same as that of the ideal FD IAB owing to the application of full SI reduction. The observed performance gap between the ideal FD IAB and the fibered-backhaul is the performance loss due to the DLI introduced earlier. While the system performance can be improved through better SI reduction, there is no straightforward method at the link-level to mitigate the DLI after the backhaul link is set as fixed; this implies that DLI should be considered at the cell-planning stage.

\begin{figure*}[t]
	\centerline{\resizebox{1.9\columnwidth}{!}{\includegraphics{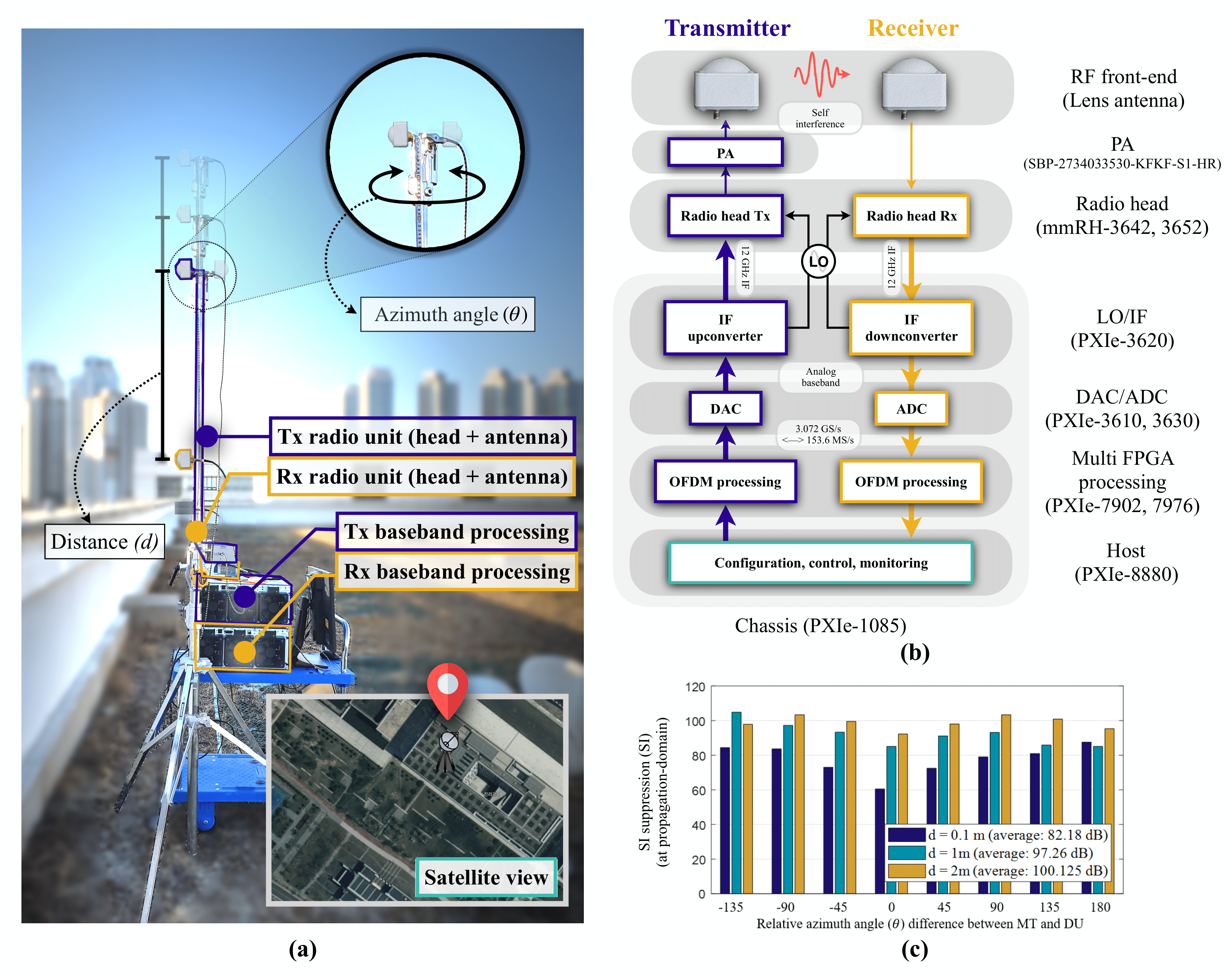}}}
	\caption{(a) Set-up of 28~GHz H/W real-time prototype in the test scenario and its satellite view; (b) Block diagram of the prototype; (c) Measured SI suppression in the propagation-domain.}	
	\label{Fig.prototype}
\end{figure*}

\section{Prototype Measurements of SI Suppression}
\label{sec.4P_Performance}

Prior observation implies that the SI suppression in the propagation-domain accounts for a considerable proportion of the total SI reduction. Based on this, we implement a \SI{28}{GHz} H/W prototype to measure the propagation-domain SI suppression \treview{with the} \SI{800}{MHz} \treview{bandwidth} (double the maximum channel bandwidth of mmWave band in 5G NR) to validate the feasibility of FD IAB. 

The prototype is implemented using the LabVIEW system design software and a PXIe software-defined radio platform based on field-programmable-gate-array (FPGA). As shown in Fig.~\ref{Fig.prototype}(a), the measurement was conducted on the outdoor rooftop of a four-floor building. The objective of the prototype was to imitate a FD IAB-node in a suburban downlink scenario with its DU and MT separated by a certain distance and having different relative orientations. The Rx below the Tx mimics the MT of an IAB-node, while the Tx plays the role of the DU. The SI is the access link's signal, which is not intended to consider the MT as an Rx (see Fig.~\ref{fig_system_model}(b)). Nevertheless, we implemented a synchronized communication link between the DU and MT to measure the received signal strength as SI. \treview{Although our measurement provides a good reference for the SI suppression performance of FD IAB by mimicking an IAB-node and its environment, optimizations in the equipment's form factor and its surroundings, which act as reflectors, are yet to be implemented. This will lead to  more enhanced and stable suppression performance.}

The block diagram in Fig.~\ref{Fig.prototype}(b) illustrates the main components of our 28~GHz prototype, where Tx and Rx are both equipped with a single antenna and an RF chain. To form a sharp analog beam at the RF front-end, advanced lens antennas with a gain of 19.86~dBi and an approximate \SI{3}{dB} beamwidth of 13.4$^{\circ}$ are used~\cite{cho2018rf}. We use an external PA with \SI{35}{dB} gain and \SI{30}{dBm} P1dB. The radio heads of the Tx/Rx up-convert/down-convert the input frequency to the target frequency, whereas the LO/IF module generates the LO and control signals for the radio heads. The DAC/ADC module is equipped with 14-bit DAC and 12-bit ADC resolutions, respectively. The FPGA processing module provides baseband OFDM processing, channel estimation, time synchronization, and digital down/up-conversion. 

Fig.~\ref{Fig.prototype}(c) illustrates the measured propagation-domain SI suppression (transmitted power $-$ received power) based on the relative azimuth angle, for various antenna separation distances. 
The average of the \SI{100.125}{dB}, \SI{97.26}{dB}, and \SI{82.18}{dB} suppressions were measured for the $d=\SI{2}{\meter}$, $d=\SI{1}{\meter}$, and $d=\SI{0.1}{\meter}$ cases, respectively, owing to the path loss and antenna directionality. Note that although they are intended to be designed similarly, the environment of the measurement and that of the simulation in Section~\ref{sec.3Performance} are not identical. Thus, the measurement and simulation results cannot be compared directly. The main differences between the two environments are the number of surrounding buildings (which act as reflectors) and the existence of the metal pole between the transceivers (which acts as an absorber in the measurement). 
The measurement results demonstrate that a large amount of SI suppression can be effectively achieved in FD IAB, especially when compared with the measurement results in the \SI{2.4}{GHz} band~\cite{EEverett2014Passive,kim2015survey}. The research team measured, for example, an SI suppression of \SI{73.8}{dB} in the case of the \SI{0.5}{m} separation distance with the absorbing shield and cross-polarized antennas. In contrast, we measured an average suppression of \SI{82.18}{dB} for the \SI{0.1}{m} separation distance with the directional antennas only. This large reduction is attributed to the sharp radiation pattern and high path loss in the system.

\section{Further Challenges and Conclusion}
\label{sec.conclusion}
This study focused on one of the most critical challenges associated with the FD IAB framework, SI reduction. To exploit the full potential of FD IAB, however, engineers must overcome further challenges that exist beyond the PHY layer. These challenges include the following: 
\begin{itemize}
	\item MAC-layer challenges: Compared with the HD IAB framework, more devices are simultaneously communicating at a given instance in FD IAB, leading to an increase in the number of interfering sources among different devices. Thus, careful resource allocation and scheduling are necessary
	\item Network-layer challenges: There is considerable potential to improve the end-to-end delay of FD IAB systems using optimized routing algorithms and protocols. The routing algorithm may vary according to the target performance measure of the application 
\end{itemize}

In this study, we investigated the performance and feasibility of FD IAB in 5G NR. We provided a brief tutorial on the FD IAB framework and its enabling technology and presented our analyses of the performance of FD IAB in a realistic network scenario. Numerical evaluations of the SI reduction at the link-level and of the downlink throughput performance at the system-level were conducted. Finally, we obtained \SI{28}{GHz} H/W prototype measurements of the propagation-domain SI suppression to verify the feasibility of FD IAB. \treview{Based on our analyses and experimental validations, we conclude that FD IAB is a feasible and promising solution for future networks.} 




\section*{Acknowledgment}
This research was in part supported by Networks Business, Samsung Electronics and IITP funded by MSIT under grants 2016-0-00208.

\bibliographystyle{IEEEtran}
\bibliography{bibtex_YearSorted}

\begin{thebibliography}{10}
\providecommand{\url}[1]{#1}
\csname url@samestyle\endcsname
\providecommand{\newblock}{\relax}
\providecommand{\bibinfo}[2]{#2}
\providecommand{\BIBentrySTDinterwordspacing}{\spaceskip=0pt\relax}
\providecommand{\BIBentryALTinterwordstretchfactor}{4}
\providecommand{\BIBentryALTinterwordspacing}{\spaceskip=\fontdimen2\font plus
\BIBentryALTinterwordstretchfactor\fontdimen3\font minus
  \fontdimen4\font\relax}
\providecommand{\BIBforeignlanguage}[2]{{%
\expandafter\ifx\csname l@#1\endcsname\relax
\typeout{** WARNING: IEEEtran.bst: No hyphenation pattern has been}%
\typeout{** loaded for the language `#1'. Using the pattern for}%
\typeout{** the default language instead.}%
\else
\language=\csname l@#1\endcsname
\fi
#2}}
\providecommand{\BIBdecl}{\relax}
\BIBdecl

\bibitem{3GPP2018NR}
\emph{{TS} 38.300, NR and NG-RAN overall description, Release 16}, 3GPP Std.

\bibitem{Samsung2014mmWave}
W.~Roh, J.-Y. Seol, J.~Park, B.~Lee, J.~Lee, Y.~Kim, J.~Cho, K.~Cheun, and
  F.~Aryanfar, ``Millimeter-wave beamforming as an enabling technology for {5G}
  cellular communications: Theoretical feasibility and prototype results,''
  \emph{{IEEE} Commun. Mag.}, vol.~52, no.~2, pp. 106--113, Feb. 2014.

\bibitem{lopez2015towards}
D.~L{\'o}pez-P{\'e}rez, M.~Ding, H.~Claussen, and A.~H. Jafari, ``Towards 1
  {G}bps/{UE} in cellular systems: Understanding ultra-dense small cell
  deployments,'' \emph{{IEEE} Commun. Surveys Tuts.}, vol.~17, no.~4, pp.
  2078--2101, Jun. 2015.

\bibitem{polese2020integrated}
M.~Polese, M.~Giordani, T.~Zugno, A.~Roy, S.~Goyal, D.~Castor, and M.~Zorzi,
  ``Integrated access and backhaul in 5{G} mmwave networks: Potential and
  challenges,'' \emph{{IEEE} Commun. Mag.}, vol.~58, no.~3, pp. 62--68, Mar.
  2020.

\bibitem{3GPP2018SI}
\emph{{TS} 38.174, Integrated access and backhaul radio transmission and
  reception, Release 16}, 3GPP Std.

\bibitem{Sachin2013Full}
D.~Bharadia, E.~McMilin, and S.~Katti, ``Full duplex radios,'' in \emph{Proc.
  Asilomar Conf. on Signal, Syst. and Comput.}, Aug. 2013, pp. 375--386.

\bibitem{Sabharwal2014IBFD}
A.~Sabharwal, P.~Schniter, D.~Guo, D.~W. Bliss, S.~Rangarajan, and R.~Wichman,
  ``In-band full-duplex wireless: Challenges and opportunities,'' \emph{{IEEE}
  J. Sel. Areas Commun.}, vol.~32, no.~9, pp. 1637--1652, Jun. 2014.

\bibitem{GLiu2015FDRelay}
G.~Liu, F.~R. Yu, H.~Ji, V.~C. Leung, and X.~Li, ``In-band full-duplex
  relaying: A survey, research issues and challenges,'' \emph{{IEEE} Commun.
  Surveys Tuts.}, vol.~17, no.~2, pp. 500--524, Jan. 2015.

\bibitem{kim2015survey}
D.~Kim, H.~Lee, and D.~Hong, ``A survey of in-band full-duplex transmission:
  From the perspective of {PHY} and {MAC} layers,'' \emph{{IEEE} Commun.
  Surveys Tuts.}, vol.~17, no.~4, pp. 2017--2046, Feb. 2015.

\bibitem{EEverett2014Passive}
E.~Everett, A.~Sahai, and A.~Sabharwal, ``Passive self-interference suppression
  for full-duplex infrastructure nodes,'' \emph{{IEEE} Trans. Wireless
  Commun.}, vol.~13, no.~2, pp. 680--694, Jan. 2014.

\bibitem{prototype}
M.~Chung, M.~S. Sim, J.~Kim, D.~K. Kim, and C.-B. Chae, ``Prototyping real-time
  full duplex radios,'' \emph{{IEEE} Commun. Mag.}, vol.~53, no.~9, pp. 56--63,
  Sep. 2015.

\bibitem{roberts2021millimeter}
I.~P. Roberts, J.~G. Andrews, H.~B. Jain, and S.~Vishwanath, ``Millimeter-wave
  full duplex radios: New challenges and techniques,'' \emph{{IEEE} Wireless
  Commun. Mag.}, vol.~28, no.~1, pp. 36--43, Feb. 2021.

\bibitem{kwack}
J.~W. Kwak, M.~S. Sim, I.-W. Kang, J.~S. Park, J.~Park, and C.-B. Chae, ``A
  comparative study of analog/digital self-interference cancellation for full
  duplex radios,'' in \emph{Proc. Asilomar Conf. on Signal, Syst. and Comput.},
  Nov. 2019, pp. 1114--1119.

\bibitem{MS_schedule}
M.~S. Sim, M.~Chung, J.~Chung, D.~K. Kim, and C.-B. Chae, ``Nonlinear
  self-interference cancellation for full-duplex radios: from link-level and
  system-level performance perspectives,'' \emph{{IEEE} Commun. Mag.}, vol.~55,
  no.~9, pp. 158--167, Sep. 2017.

\bibitem{cho2018rf}
Y.~J. Cho, G.-Y. Suk, B.~Kim, D.~K. Kim, and C.-B. Chae, ``{RF} lens-embedded
  antenna array for mm{W}ave {MIMO}: Design and performance,'' \emph{{IEEE}
  Commun. Mag.}, vol.~56, no.~7, pp. 42--48, Jul. 2018.

\end{thebibliography}

\vspace{-6pt}
\end{document}